# MaPBI3 and 2D hybrid organic inorganic perovskite based microcavities employing periodic, aperiodic and disordered photonic structures and with the possibility of light-induced tuning

Michele Bellingeri[1,2], Francesco Scotognella[2,3]*

[1] Dipartimento di Fisica, Politecnico di Milano, Piazza Leonardo da Vinci 32, 20133, Milano, Italy

[2] Dipartimento di Scienze Matematiche, Fisiche e Informatiche, Università di Parma, via G.P. Usberti, 7/a, 43124 Parma, Italy

[3] Center for Nano Science and Technology@PoliMi, Istituto Italiano di Tecnologia, Via Giovanni Pascoli, 70/3, 20133, Milan, Italy

* E-mail address: francesco.scotognella@polimi.it

**Abstract**

Inorganic-organic perovskites semiconductors are becoming increasingly interesting due to their remarkable optical properties, such as a high photoluminescence quantum yield and the possibility to show optical gain in a broad range of wavelengths. We have here simulated microcavities that embed MaPBI3 and 2D hybrid organic inorganic perovskite semiconductors by sandwiching such active layers between periodic, aperiodic and disordered photonic structures. We have carefully considered the refractive index dispersion of all the materials employed, such silicon dioxide, titanium dioxide and the perovskite layers. Moreover, by employing a photochromatic polymer, namely the diarylethene-based polyester pDTE, we have designed a microcavity with light-induced tuning of the cavity modes is possible.

**Introduction**

Organic-inorganic halide perovskite semiconductors have become in few years a main player in photonics due to their impressive nonlinear optical properties [1–3]. These materials are interesting in the materials science community because of their unique performances as photovoltaic materials, but also because of their high photoluminescence quantum yield and optical gain, making them promising for light emitting diodes and lasers [4]. Just to cite few examples, perovskite-based lasers have been monolithically integrated into a silicon nitride photonic circuit [5] and perovskite-based random lasing with cavity exciton-polaritons has been reported [6].

In photonics one of the most simple building blocks is the photonic crystal [7–9], characterized by the alternation of different refractive indexes in one, two or three directions. The one-dimensional photonic crystals are particularly interesting because they can be fabricated with a large variety of techniques [10], the optical properties can be studied with high accuracy [11], for enhancement of spontaneous emission [12,13] and lasers [14–18]. One-dimensional photonic structures can be designed by following aperiodic and disordered (random) patterns. These structures can be exploited to tune and enhance the optical characteristics of emitting materials embedded in such structures [19–22].

One-dimensional photonic crystals have been with perovskites to fabricate efficient solar cells with tunable structural colour [23]. Moreover, microcavities based on polymeric materials that embed perovskites have been designed and fabricated [24].

For a proper design of the photonic structures that include perovskite semiconductors as a constituent material, it is beneficial to know the real part and the imaginary part of the refractive index as a function of the wavelength, i.e. the complex refractive index dispersion. The refractive index of MAPbI3 has been reported by several groups in Refs. [25,26] and [27,28]. Recently, also the refractive indexes of 2D hybrid organic inorganic perovskite (HOIP) semiconductors in Ruddlesden–Popper (RP) and Dion–Jacobson (DJ) phases have been reported [29].

In this work we exploit the determination of the complex refractive index dispersion of such materials and we have simulated microcavities that embed MaPBI3 and 2D hybrid organic inorganic perovskite semiconductors by sandwiching such active layers between periodic, aperiodic and disordered photonic structures. We demonstrate the possibility to observe cavity modes with the different structures. Moreover, we demonstrate the possibility to obtain a tunable microcavity by adding as layers in the microcavity a diarylethene-based polyester (pDTE).

**Methods**

We have used the transfer matrix method [30–32] to simulate the light transmission spectra of the different designed microcavities.

In this study we consider a system glass/multilayer/air that is impinged by the light with normal incidence. The multilayer system can be described with a product of matrix $\prod_{j=1}^{x} M_j = M = \begin{bmatrix} m_{11} & m_{12} \\ m_{21} & m_{22} \end{bmatrix}$ with $x$ number of layers. The characteristic matrix of each layer $M_j$, with $j=(1,2,…,x)$, is:

$$M_j = \begin{bmatrix} \cos(\phi_j) & -\frac{i}{p_j}\sin(\phi_j) \\ -ip_j\sin(\phi_j) & \cos(\phi_j) \end{bmatrix} \quad (1)$$

$\phi_j$, argument of the trigonometric functions in the matrix, is the phase variation of the light wave passing through the $j$th layer, and for normal incidence $\phi_j = (2\pi/\lambda)n_j d_j$, where $n_j$ is the refractive index of the layer and $d_j$ its thickness. The parameter $p_j = \sqrt{\varepsilon_j/\mu_j}$ in transverse electric (TE) wave, while $q_j=1/p_j$ replace $p_j$ in transverse magnetic (TM) wave (taking into account that, at normal incidence, the transmission spectra for TE and TM waves are the same).

From the final matrix $M$ we can determine the transmission coefficient

$$t = \frac{2p_s}{(m_{11}+m_{12}p_0)p_s+(m_{21}+m_{22}p_0)} \quad (2)$$

And, consequently, the transmission:

$$T = \frac{p_0}{p_s}|t|^2 \tag{3}$$

The dispersion of the refractive index for SiO$_2$ has been taken from Ref. [33]:

$$n^2_{SiO_2}(\lambda) - 1 = \frac{1.9558\lambda^2}{\lambda^2 - 0.15494^2} + \frac{1.345\lambda^2}{\lambda^2 - 0.0634^2} + \frac{10.41\lambda^2}{\lambda^2 - 27.12^2} \tag{4}$$

While the dispersion of the refractive index of TiO$_2$ has been taken from Ref. [34]:

$$n_{TiO_2}(\lambda) = \left(4.99 + \frac{1}{96.6\lambda^{1.1}} + \frac{1}{4.60\lambda^{1.95}}\right)^{1/2} \tag{5}$$

The complex refractive index dispersion has been taken from Refs. [27,28], while the refractive index dispersions for two-dimensional hybrid organic-inorganic perovskite semiconductors (Ruddlesden-Popper (RP) phase and Dion-Jacobson (DJ) phase) have been taken from Ref. [29].

**Results and Discussion**

We have designed three different microcavities, characterized by perovskite layers embedded between two periodic photonic crystals, two aperiodic photonic crystals, or two disordered photonic structures. We have employed three different types of perovskite semiconductors: MAPbI$_3$, monolayered Ruddlesden-Popper (RP) phase and monolayered Dion-Jacobson (DJ) phase.

The periodic microcavity that embeds the perovskite layer consists of a photonic crystal of 10 bilayers of TiO$_2$ and SiO$_2$, the layer of MAPbI$_3$, another photonic crystal of 10 bilayers of TiO$_2$ and SiO$_2$. The aperiodic microcavity consist of two Thue-Morse quasicrystals that embed the perovskite layer. Each Thue-Morse crystal is composed by 32 layers with the following sequence: ABBABAABBAABABBABAABABBAABBABAAB, in which A is TiO$_2$ and B is SiO$_2$ [35]. The disordered microcavity consists of two disordered photonic structures that embeds the perovskite layer. Each disordered structure is composed by 32 layers in which each layer has a 50% probability of be either TiO$_2$ or SiO$_2$. The sequence of layers for the first disordered structure is BBABBAABBBABBABAABBBBABBBBBABABA, while the sequence for the second structure is AAABBABAAABBAAABBBABBAAABABABABB.

For the MAPbI$_3$-based microcavities, the thicknesses of the layers are 135 nm, 175 nm, and 350 nm, for TiO$_2$, SiO$_2$, and MAPbI$_3$, respectively. In Figure 1 we show the transmission spectra of microcavities embedding MAPbI$_3$, with the periodic microcavity in Figure 1 top, the aperiodic Thue-Morse microcavity in the center, and the disordered microcavity in Figure 1 bottom.

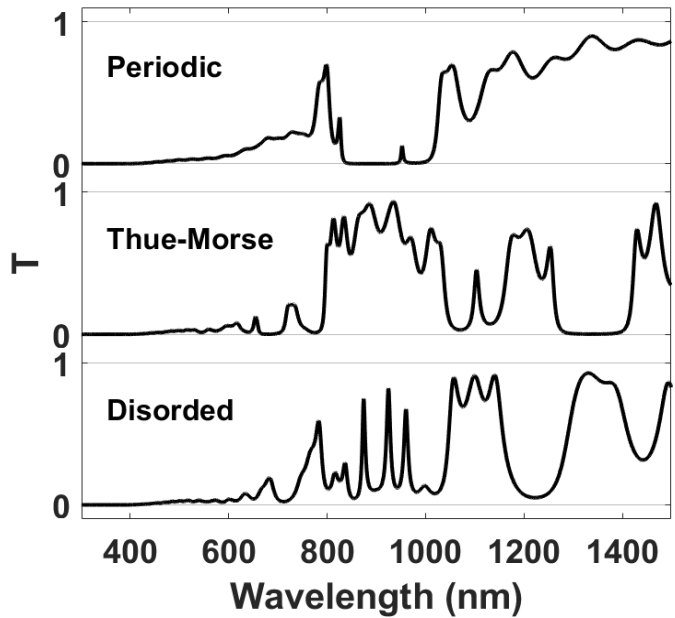

Figure 1. MAPbI3 layer embedded in periodic, aperiodic (Thue-Morse), disordered microcavities.

We have simulated microcavities in which the embedded layer is the Ruddlesden-Popper (RP) phase of a two-dimensional hybrid organic-inorganic perovskite semiconductor. We have selected the monolayer of the perovskite semiconductor, that we call it RP1 as in Reference [29]. In the case of the monolayer, the formula of the compound is $(CH_3(CH_2)_3NH_3)_2PbI_2$ [29]. In Figure 2 we show the transmission spectra of a periodic (top), an aperiodic Thue-Morse (center) and a disorder cavity (bottom). For these cavities the thickness of the titanium dioxide layers is 93 nm, the thickness of the silicon dioxide layers is 120 nm, and the thickness of the RP perovskite layer is 350 nm.

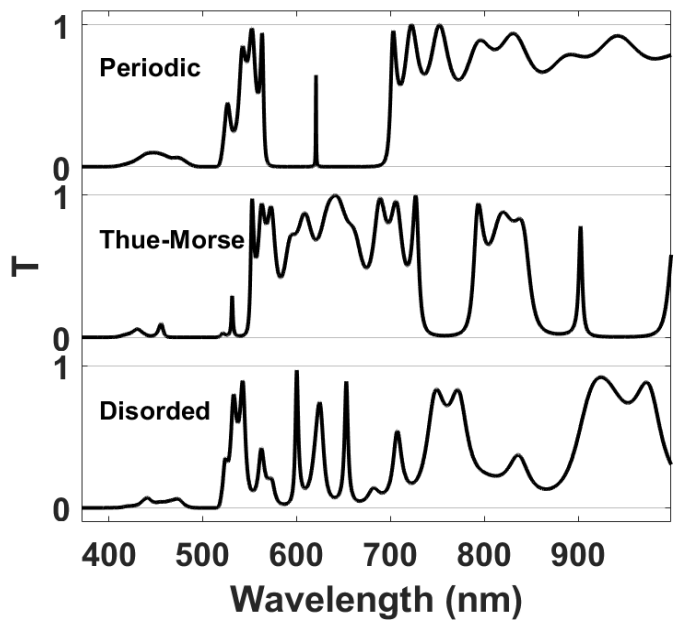

Figure 2. RP1 layer embedded in periodic, aperiodic (Thue-Morse), disordered microcavities.

Finally, we have simulated microcavities in which the embedded layer is the Dion-Jacobson (DJ) phase of a two-dimensional hybrid organic-inorganic perovskite semiconductor. We have selected the monolayer of the perovskite semiconductor, that we call it DJ1 as in Reference [29]. In the case of the monolayer, the formula of the compound is 4-(aminomethyl)piperidinium $PbI_4$ [29]. In Figure 3 the transmission spectra of the DJ1 phase embedded in a periodic (top), an aperiodic cavity (center) and a disordered cavity (bottom). The thickness of the layers for these structures with DJ1 is 120 nm, 150 nm and 350 nm for $TiO_2$, $SiO_2$ and DJ1, respectively.

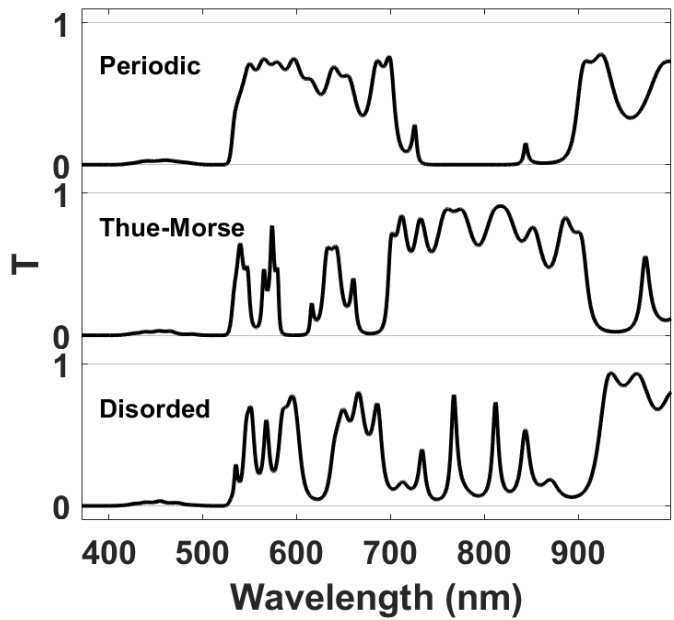

Figure 3. DJ1 layer embedded in periodic, aperiodic (Thue-Morse), disordered microcavities.

To design a tunable structure we have employed a photochromic polymer, pDTE [36], as additional layer to induce with light a shift of the spectral feature of the microcavity. It has been reported that ultraviolet irradiation induces a change from a colourless form of pDTE to a blue form of pDTE [36]. In this case, the microcavity is composed by a 10-bilayer $TiO_2$-$SiO_2$ photonic crystal, a layer of pDTE, a layer with the monolayered RP phase, a layer of pDTE, and another 10-bilayer $TiO_2$-$SiO_2$ photonic crystal. In Figure 4 we show the transmission spectra of the cavity in the case of the transparent phase of pDTE (solid curve) and in the case of the blue phase of pDTE (dashed curve).

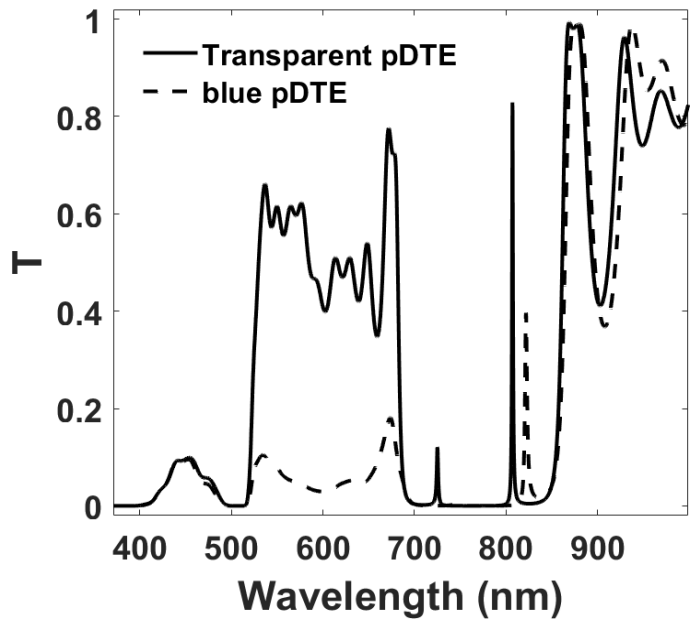

Figure 4. Periodic cavity [(SiO$_2$/TiO$_2$)x10]/pDTE/PR1/pDTE/[(SiO$_2$/TiO$_2$)x10] with transparent phase of pDTE (solid curve) and blue phase of pDTE (dashed curve).

The complex refractive index of pDTE in the two forms has been taken from Reference [36]. The transmission valley centred at 600 nm for the blue form is due to the strong absorption of the form in this spectral region. We can observe a shift of 15 nm in the cavity mode at longer wavelengths (about 800 nm) due to the photochromic behaviour of pDTE.

**Conclusion**

To conclude, we have designed by using the transfer matrix method one-dimensional microcavity that embed MAPbI$_3$ and 2D hybrid organic inorganic perovskite semiconductors. Such active materials have been sandwiched between periodic, aperiodic and disordered photonic structures. To engineer the perovskite layer we have taken in to account the complex refractive index dispersion of the materials, together with the refractive index dispersions of silicon dioxide and titanium dioxide. In order to conceive a tunable perovskite based microcavity, we have employed pDTE, which is photochromic materials able to change the refractive index upon ultraviolet irradiation.